# Implementation of a planar coil of wires as a sinus-galvanometer. Analysis of the coil magnetic field

**Dimitar G Stoyanov**
Sofia Technical University, Sliven Engineering and Pedagogical Faculty,
59 Burgasko Shosse Blvd, 8800 Sliven, Bulgaria

E-mail: dgstoyanov@abv.bg

**Abstract**
The paper presents a theoretical analysis on the interaction between the Earth's magnetic field of a compass needle and the magnetic field of a straight infinite current-carrying wire. Implementation of a planar horizontal coil of wires has been shown as a sinus galvanometer. The magnetic field over the planar coil of wires has been examined by experiment. The coil could be used as a model for straight infinite current wire in demonstration set-ups or could be given as an assignment in Physics laboratory workshops.

## 1. Introduction
In 1820, Christian Oersted noticed that a compass needle deflected in the presence of a current-carrying wire. However, the classical set-up employs a source of steady power supply and a current of 10-20 A, which makes the demonstration very difficult.

Hence, our aim is to modify the Oersted experiment to make it easy and low-cost and to use simple voltage sources. The implementation of a planar coil of wires, which deflects vigorously the compass magnetic needle (more than 80 angular degrees) when a current of up to 1 A flows along it, has been reported. (See [1]). The planar coil creates a magnetic field in the middle of the bundle of wires , which is very close to the magnetic field of a single infinite straight current-carrying wire. In [1] theoretically presents the interaction between the magnetic needle of a compass and the magnetic field created by the coil. As a result, the possibility for the coil and the compass to be used as a tangent galvanometer has been theoretically and experimentally proved. Strong nonlinearities in the functional dependences have been ascertained taking into account the inhomogeneity of the magnetic field around the current-carrying wire.

Thus, the need to find another way to create a galvanometer using this horizontal coil emerges. The aim of the following paper is to point out the possibility to use the coil and the compass as a sinus galvanometer.

## 2. Theoretical description of the interaction between a compass and a current-carrying wire
### 2.1 *A magnetic needle in the magnetic field of a wire*
The theoretical analysis of the interaction between the magnetic field of a wire and the magnetic needle will be carried out using the approach in [1]. First, the interaction between a magnetic needle and the magnetic field of a straight infinite horizontal current-carrying wire will be examined.

Let a current-carrying wire lie on the **OY** -axis of the Cartesian coordinate system **K** (figure 1). A steady current flows along the wire with a magnitude of **I**, having the same positive direction as **OY**- axis. At a point on the **OZ**-axis the suspension point lies at a distance of **z** from the origin of the coordinate system of a magnetic needle of a length $\Delta$. The pivot of the magnetic needle coincides with the **OZ**-axis while the magnetic needle lies and moves on a plane parallel to **XOY** plane. The actual position of the magnetic needle is characterized by the rotation angle of the needle $\theta$ with respect to the positive direction of **OY**-axis (figure 1). The magnetic needle has a magnetic dipole moment of **m**.

The flowing current creates a magnetic field around the wire. The magnitude of the horizontal component $\mathbf{B_I}$ of the magnetic induction vector in the positive direction of **OX**-axis is [1]:

$$\mathbf{B_I} = \frac{\mu_0 . I}{2.\pi} . \frac{z}{\rho^2} \qquad (1)$$

where $$\rho = \sqrt{z^2 + x^2}. \qquad (2)$$

is the distance from the given point with coordinates of (**x, y, z**) to the wire axis.

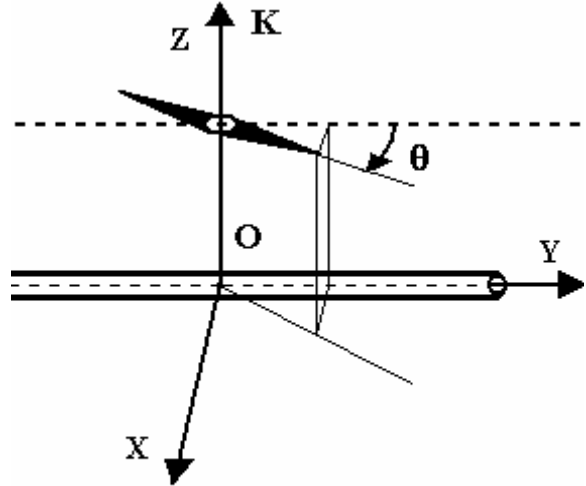

**Figure 1. Mutual disposition of a magnetic needle and a horizontal wire.**

Due to the inhomogeneity of the magnetic field around the wire the torque of the magnetic needle $\mathbf{M_I}$ is obtained by the integration of the infinite small torques created by the different infinite small parts of the magnetic needle.

The torque of the whole needle, $\mathbf{M_I}$, is [1]:

$$\mathbf{M_I} = -\frac{\mu_0 . I}{2.\pi.z} . m . \cos\theta . \int_0^\Delta \frac{z^2}{\rho^2} . \frac{dm}{m} = -\frac{\mu_0 . I}{2.\pi.z} . m . \cos\theta . f(\delta) \qquad (3)$$

The integral on the right side of (3) is a dimensionless function $\mathbf{f(\delta)}$ of the parameter $\delta$ (which is also dimensionless):

$$\delta = \frac{\Delta . \sin\theta}{2.z}. \qquad (4)$$

Generally, the function of **f(δ)** is even and decreases with an increase of δ due to the needle symmetry which takes the pivot into account. At zero, the function of **f(δ)** has a value of 1.

## 2.2 *A magnetic needle in the magnetic field of the Earth*

The horizontal component of the induction of Earth's magnetic field is represented by $\vec{B}_e$. If vector $\vec{B}_e$ is at angle β with the **OY** – axis of the coordinate system, it has the following components in the same coordinate system

$$\vec{B}_e = (-B_e \sin\beta,\ B_e \cos\beta,\ 0). \tag{5}$$

The geomagnetic field is a homogeneous one. The interaction of the magnetic needle with it results in a torque of $M_e$:

$$M_e = -m \cdot B_e \cdot \sin(\theta - \beta). \tag{6}$$

## 2.3 *A magnetic needle in the Earth's and the wire's magnetic fields*

We assume the magnetic needle has a moment of inertia, **J**, towards its pivot. Taking into consideration (3) and (6), the simultaneous action of both the Earth's magnetic field and the one created by the wire can be expressed by following equation of twisting motion [1]:

$$J \cdot \frac{d^2\theta}{dt^2} = -m \cdot B_e \cdot \sin(\theta - \beta) - m \cdot \frac{\mu_0 \cdot I}{2 \cdot \pi \cdot z} \cdot \cos\theta \cdot f(\delta). \tag{7}$$

The above equation has the following stationary solutions.
(i) At zero current along the wire and applying (7), we obtain: $\theta = \beta$, and the needle orients itself along the Earth's magnetic field.
(ii) At **θ = 0** all parts of the magnetic needle are at the same distance from the wire, and **δ = 0**. Applying (7), we obtain:

$$B_e \cdot \sin\beta = B_I = \frac{\mu_0 \cdot I}{2 \cdot \pi \cdot z}. \tag{8}$$

After transformation we obtain:

$$I = \frac{2 \cdot \pi \cdot z \cdot B_e}{\mu_0} \cdot \sin\beta = I_e \cdot \sin\beta. \tag{9}$$

where $I_e$ is the equivalent to the current of Earth's magnetic field when **z** is given:

$$I_e = \frac{2 \cdot \pi \cdot z \cdot B_e}{\mu_0}. \tag{10}$$

Equation (9) can be used as a method for measuring the current magnitude, which could be called a sinus galvanometer.

## 3. Current coil in the mode of sinus galvanometer

So far, the theoretical analysis to use the compass magnetic needle as a sinus galvanometer has been presented. To implement it in reality would require greater accuracy than as a tangent galvanometer.

Moreover, the measured current values must be limited to the value of $I_e$ (taking into account (9) $|I| \leq I_e$)

*3.1 A measuring procedure for a sinus galvanometer*

The procedure to measure the current along the wire is as follows:

1. The noncurrent-carrying wire orients itself along the north-south direction by the compass (figure 2a). The compass scale is set to zero. The wire and the compass are steadily fixed, respectively.

2. A current flows along the wire with a magnitude which is of interest to us (figure 2b). The magnetic needle deflects at a certain angle.

3. The current-carrying wire and the compass are rotated simultaneously as far as the compass needle sets to zero (figure 2c).

4. The current along the wire is switched off and the deflected angle of $\beta$ is determined by the compass scale (figure 2d). The current magnitude is estimated applying (9).

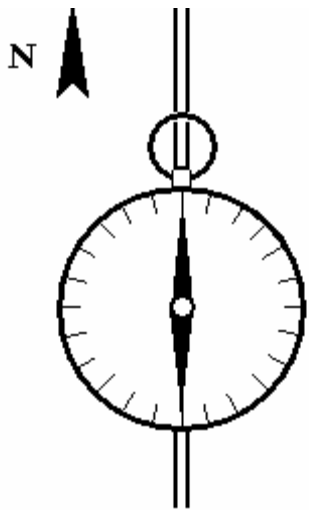

**Figure 2a. Initial position**

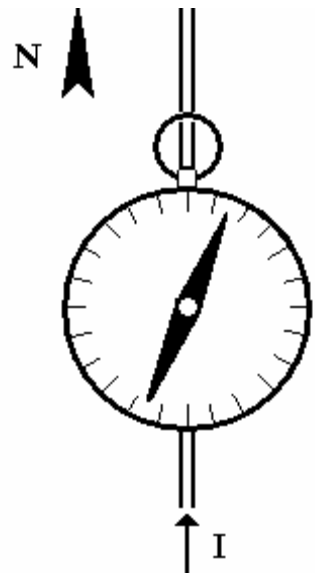

**Figure 2b. Current is on**

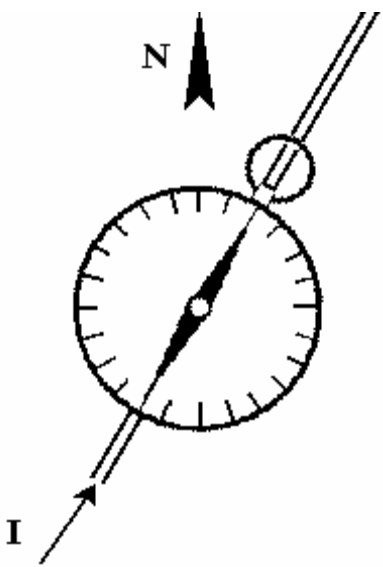

**Figure 2c. Rotation of the wire**

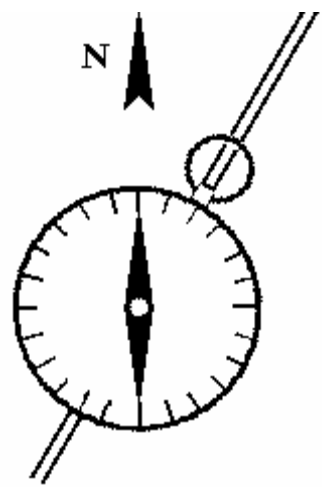

**Figure 2d. Current is off**

*3.2 Description of the set-up and consideration of results*

The above-mentioned set-up was carried out using a planar horizontal coil of wires described in [1] and shown in figure 3. The total number of windings in the bunch was 24 (**N=24**). When a current of **I** flows along each current-carrying wire, the total current along the bunch windings is **I.N**. A steady current source was provided by a dc power supply rectifier with an output voltage of up to 15 V and a max current of 2 A. A resistance of 15 $\Omega$ with a dispersing power of 20 W was connected in serial as a ballast into the planar coil.

In the set-up, the planar coil was placed horizontally and the compass lay in the middle of the bunch windings (See figure 3).

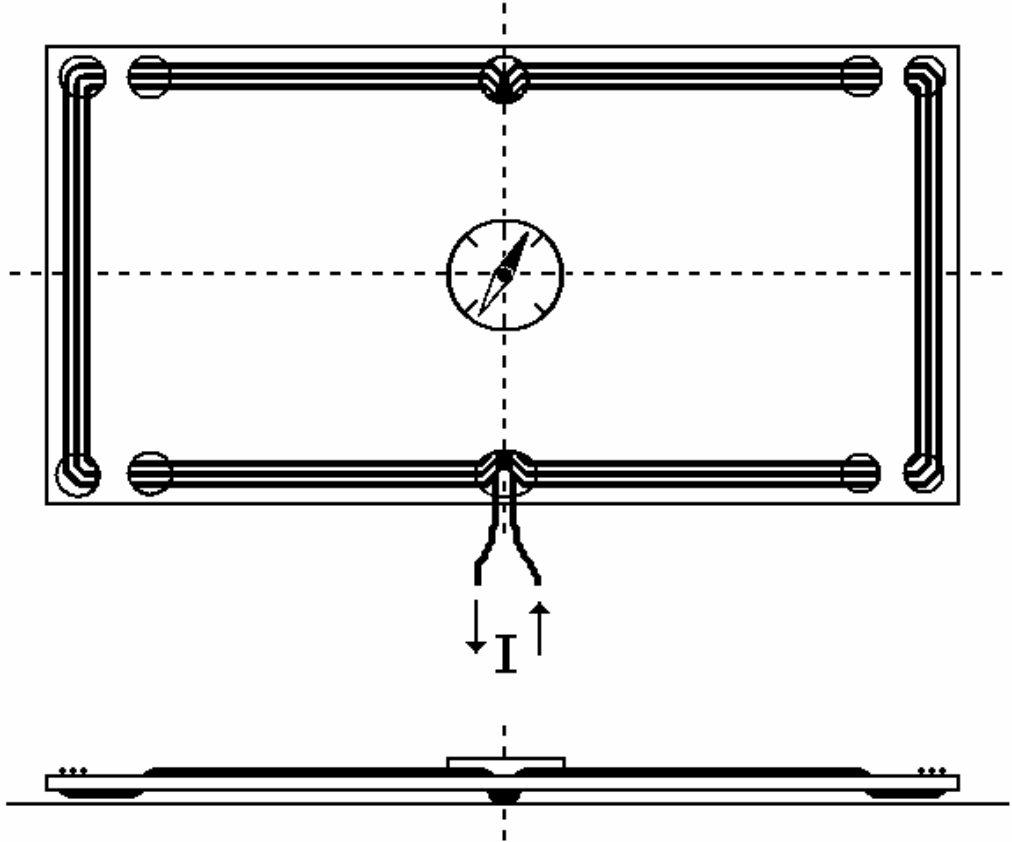

**Figure 3. Framework of the planar coil of wires used in the set-up**

First, the bunch windings were oriented along the north-south direction by the compass as the current was switched off along the frame. The compass scale was set to zero. Thus, the magnetic field created by the bunch windings, $\mathbf{B}_I$, had a direction, which was perpendicular to the direction of the horizontal component of the Earth's magnetic field, $\mathbf{B}_e$.

Then, the current was switched on along the frame and the magnetic needle deflected at an angle of $\theta$. The angle of deflection on the left has a positive sign (See figure 2b).

Next, the frame and the compass and the compass on it were rotated so as to reset the compass needle to zero (figure 2c).

Finally, the current was switched off and the deflection angle, $\beta$, was determined by the compass scale.

Taking the experiment into account and applying (9), we obtain

$$I.N = \frac{2.\pi.B_e}{\mu_0}.z.\sin\beta \qquad (11)$$

The range of the deflection angle, $\beta$, was from $-75^o$ to $+75^o$ at an interval of $15^o$. Figure 4 represents the measurements obtained from the multiplication of the current, **I**, by the number of windings, **N**, as a function of the deflection angle, $\beta$, of the compass needle acting as a sinus galvanometer. The graph also displays the linear dependence of the current on $sin\,\beta$ with a high coefficient of correlation.

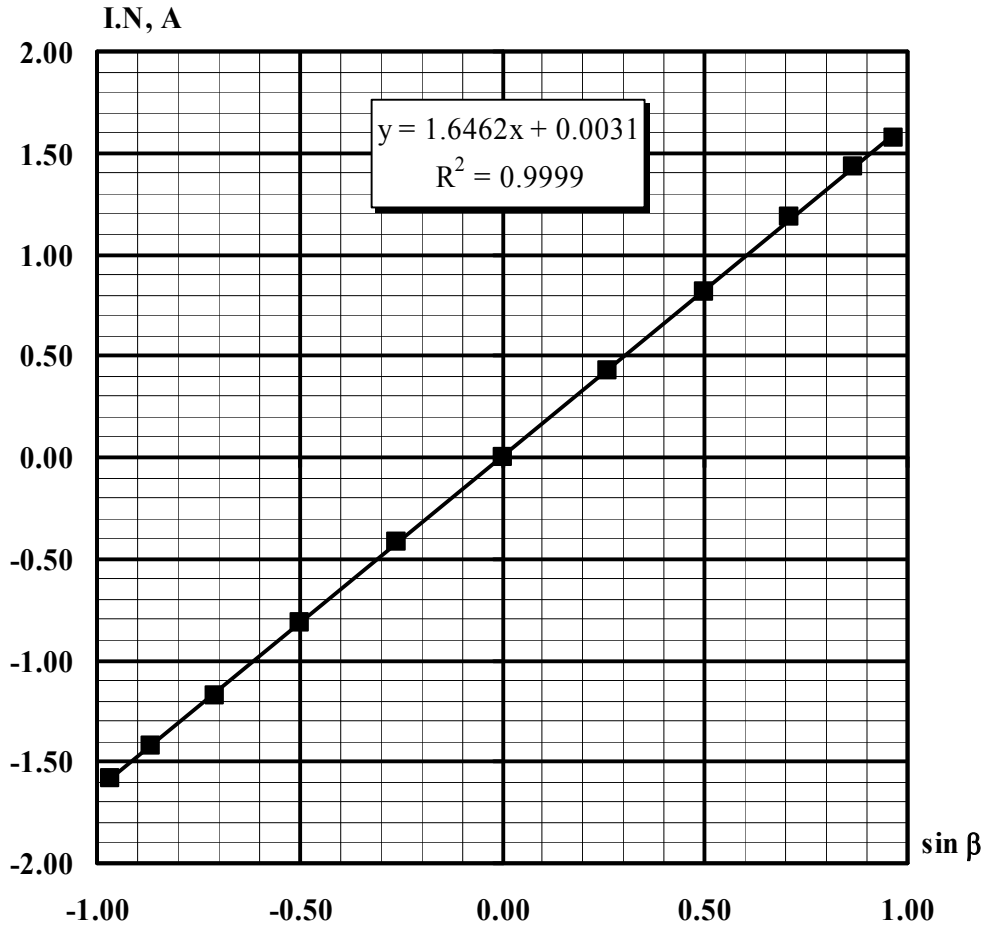

**Figure 4. Measurements taking into account (9) applying $sin\,\beta$.**

As a result, the deflected angle of the compass needle measured as precisely as possible makes us believe that the horizontal coil of wires could be used as a sinus galvanometer either for demonstrations or could be given as an assignment in Physics laboratory workshops. For instance, the obtained average value of $I_e = 1.6462\,A$ in the measurements (figure 4) and at $z = 11.9\,mm$ the magnitude of the horizontal component of Earth's magnetic field has been estimated at $B_e = (27.69 \pm 0.05)\times 10^{-6}\,T$ at Sofia Technical University, Sliven Engineering Pedagogical Faculty.

### 4. Analysis of the planar coil magnetic field

The linearity of the dependence (11) and the simple calculations urged us to use the planar coil as a sinus galvanometer in estimating the magnetic field created in the area of the bunch windings of the coil. We take the Earth's magnetic field as a standard and for the magnetic field created by the current applying (11), we obtain

$$\frac{B_I}{B_e} = \sin\beta = \frac{\mu_0.I.N}{B_e.2.\pi.z}. \tag{12}$$

Hence, the magnetic field of the coil grows in linearity with the current increase of **I.N** at one and the same distance of the compass from the wire **z**. Moreover, the inclination of the straight line is disproportional to **z**.

The distance from the wire to the compass **z** could vary by placing several pads under the compass. To make the analysis of the dependence, $B_I$, easy a transformation with the same linear connection along **z** (at constant $\beta$) could be applied:

$$\frac{I.N}{\sin\beta} = \frac{2.\pi.B_e}{\mu_0}.z. \qquad (13)$$

**4.1 *Description of the set-up and consideration of results***
In the set-up, the frame was placed horizontally on a plane and the compass lay in the middle of the bunch windings (See figure 5). A number of dielectric plates, **n = 0, 1, 2, 3, 4**, with one and the same width of **d = 8.6 mm** were put between the coil plane and the compass.

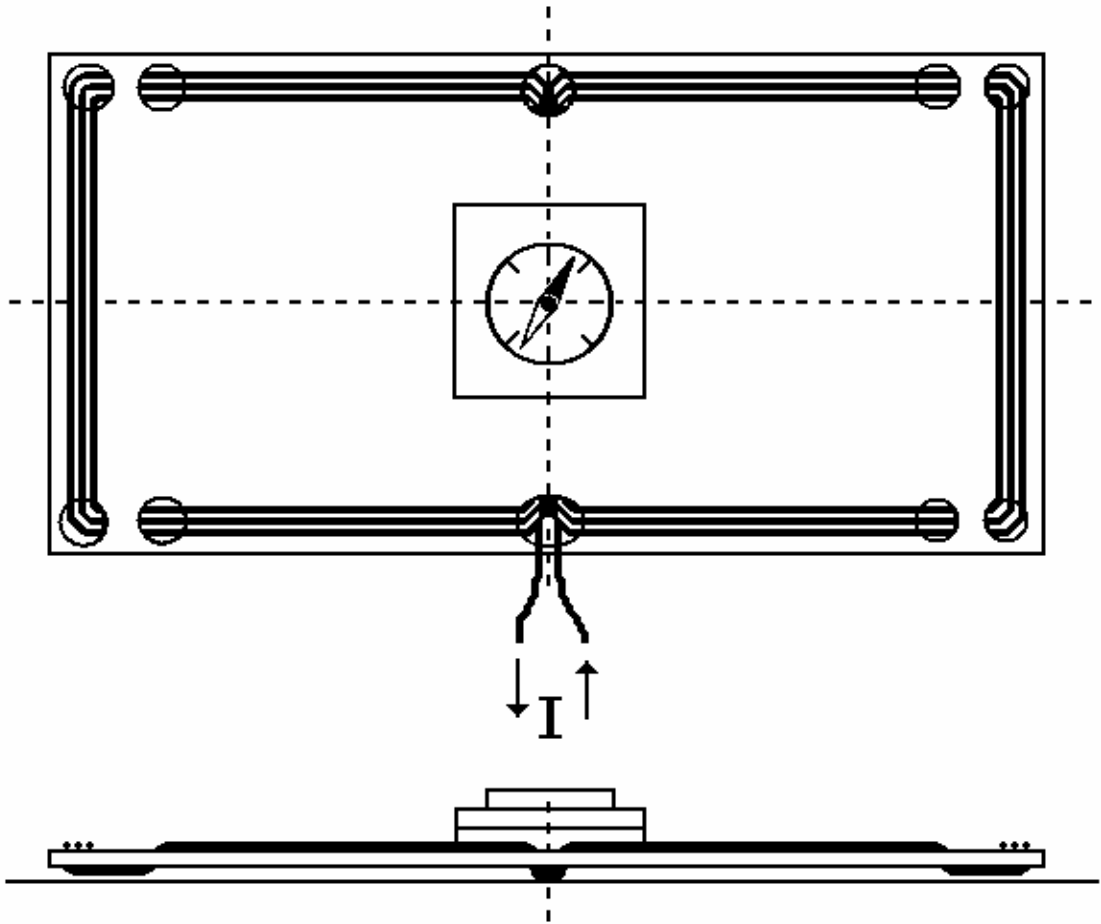

**Figure 5. Location of the compass and 2 pads**

The dielectric plates set the compass needle from the wire at a distance of

$$z = z_0 + n.d. \qquad (14)$$

Here $z_0 = 11.9\,\text{mm}$ is the minimum distance between the wire and the compass magnetic needle, which is obtained when the compass directly lies on the plastic sheet of the planar coil.

The experiment was carried out as follows:

First, the bundle windings were oriented along the north-south direction by the compass on an off-current coil of wires and the compass scale was set at zero.

Then, the current, **I**, was switched on along the frame and the magnetic needle deflected at an angle of $\theta$. The angle of deflection on the left has a positive sign (See figure 2b).

Next, the frame and the compass on it were rotated so as to reset the compass needle to zero (figure 2c).

Finally, the current was switched off and the deflection angle, $\beta$, was determined by the compass scale (figure 2d).

However, the dependences of (12) and (13) could be determined easily if the angle, $\beta$, was given to determine the current which sets the compass needle to zero, the values range of the deflection angle, $\beta$, were from $0^o$ to $+75^o$ at an interval of $15^o$.

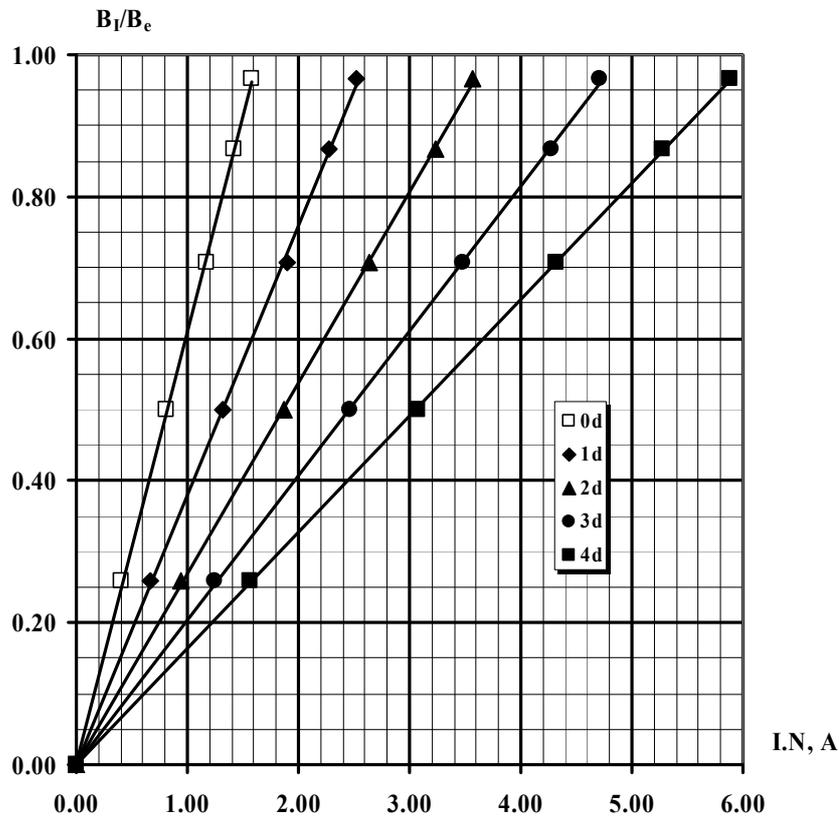

**Figure 6. Variations of the coil magnetic field as a function of the current**

Figure 6 shows the measurements of $B_I / B_e$ as a function of the multiplication of the current, **I**, by the number of windings, **N**, using different number of pads under the compass. The graph also displays a linear dependence on the current along the coil taking into consideration the measuring accuracy of the angle, $\beta$. Moreover, it could also serve as an experiment check-up of (12) confirming the linear dependence on the current.

Finally, the graph shows that the inclination of the straight line decreases with the increase of **n**, which leads to the rise of **z** (as a consequence of (14)).

*4.2 Consideration of results*
The same experimental results can be obtained using (13). Figure 7 shows the measurements of the multiplication of the current, **I**, by the number of windings, **N**, as a function of the pads number placed under the compass at one and the same magnetic needle's deflection angle, $\beta$. A linear dependence of the current on *n* can be observed as all lines cross at one and the same point of the X – axis equal to $-z_0 / d$. The obtained graph verifies (13).

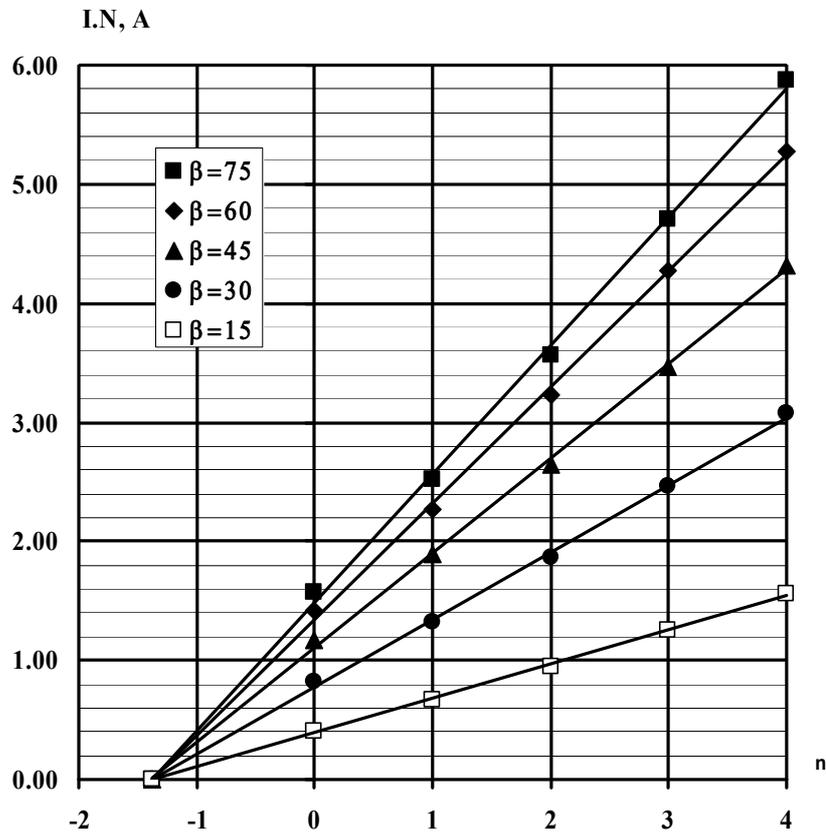

**Figure 7. Current dependence on the number of pads at one and the same angle of $\beta$**

The statistics processing of the experimental results provides the following average curve:

$$\frac{I.N}{\sin \beta} = 1.545 + 1.12n, \quad A. \tag{15}$$

The obtained average value of $I_e(z_0) = 1.545\,A$ differs with 6% from the one we have in paragraph 3.2. The variation could be explained by the scale divisions of the compass used in the set-up. Each one represents $3^0$ which makes it possible for accidental errors of up to $0.5^0$ to occur in the angle.

Figure 8 shows the theoretical dependence and the experimental data of the relation between the magnetic field $B(h)$ at the height of $h$ over the planar coil of wires $B(0)$ as a height function over it. The graph displays that the planar coil of wires magnetic field is close to the theoretical curve (8) at the distance of up to 4 cm over it. Data points correspond to $h = n.d$.

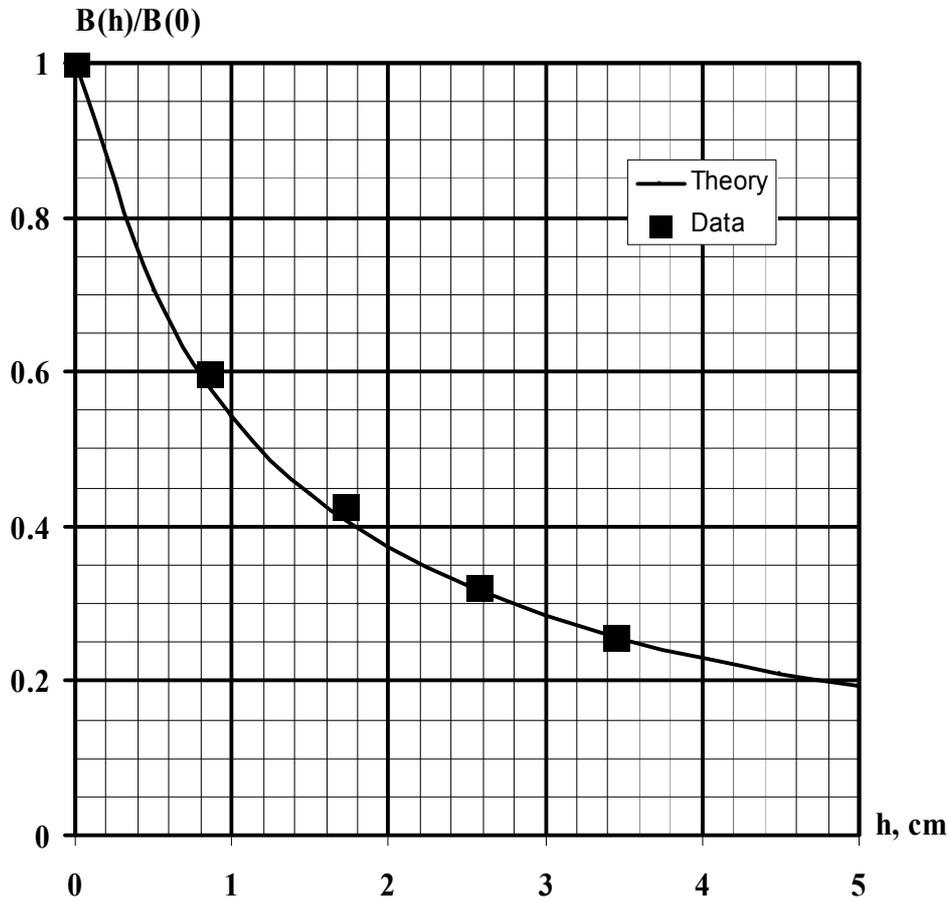

**Fig. 8 Comparison between the theoretical dependence and the experimental data**

### 5. Conclusion
To sum up, a theoretical analysis of the interaction between the compass magnetic needle and the current-carrying wire's magnetic field has been done. The possibility to use a horizontal coil of wires as a sinus-galvanometer has been shown and tested by experiment.

Therefore, we ascertain the magnetic field of the planar coil of wires is close to its theoretical value at the distance of up to 4 cm over it, taking into account the measuring accuracy of the needle's deflection angle. Hence, the planar coil of wires can be used as a model for a straight infinite current-carrying wire in demonstrations or can be given as an assignment to students in Physics laboratory workshops.